\documentclass[conference]{IEEEtran}
\IEEEoverridecommandlockouts
\usepackage{cite}
\usepackage{amsmath,amssymb,amsfonts}
\usepackage{algorithmic}
\usepackage{graphicx}
\usepackage{tabularx}
\usepackage{textcomp}
\usepackage{xcolor}
\usepackage{subcaption}
\usepackage{makecell}

\usepackage{array}
\usepackage{amsmath}
\usepackage{calc}
\usepackage{bm}
\usepackage{lipsum}
\usepackage{array,multirow,graphicx}
\usepackage[bookmarks=false]{hyperref}
\newcolumntype{P}[1]{>{\centering\arraybackslash}p{#1}}
\newcolumntype{M}[1]{>{\centering\arraybackslash}m{#1}}

\def\BibTeX{{\rm B\kern-.05em{\sc i\kern-.025em b}\kern-.08em
    T\kern-.1667em\lower.7ex\hbox{E}\kern-.125emX}}
\begin{document}
\newcommand{\RomanNumeralCaps}[1]
    {\MakeUppercase{\romannumeral #1}}
\def\BigRoman{\uppercase\expandafter{\romannumeral\number\count 255 }}
\def\Romannumeral{\afterassignment\BigRoman\count255=}

\title{Data-aided Active User Detection with a User Activity Extraction Network for Grant-free SCMA Systems}

\author{\IEEEauthorblockN{Minsig Han\IEEEauthorrefmark{1},
Ameha T. Abebe\IEEEauthorrefmark{2}, and
Chung G. Kang\IEEEauthorrefmark{1}}
\IEEEauthorblockA{\IEEEauthorrefmark{1}School of Electrical Engineering, Korea University, Seoul, Republic of Korea\\
\IEEEauthorrefmark{2}Saumsung Research, Seoul, Republic of Korea\\
Email: \IEEEauthorrefmark{1}\{als4585, ccgkang\}@korea.ac.kr,
\IEEEauthorrefmark{2}amehat.abebe@samsung.com}}
\maketitle


\begin{abstract}
In grant-free sparse code multiple access (GF-SCMA) system, active user detection (AUD) is a major performance bottleneck as it involves complex combinatorial problem, which makes joint design of contention resources for users and AUD at the receiver a crucial but a challenging problem. To this end, we propose autoencoder (AE)-based joint optimization of both preamble generation networks (PGNs) in the encoder side and data-aided AUD in the decoder side. The core architecture of the proposed AE is a novel user activity extraction network (UAEN) in the decoder that extracts a priori user activity information from the SCMA codeword data for the data-aided AUD. An end-to-end training of the proposed AE enables joint optimization of the contention resources, i.e., preamble sequences, each associated with one of the codebooks, and extraction of user activity information from both preamble and SCMA-based data transmission. Furthermore, we propose a self-supervised pre-training scheme for the UAEN prior to the end-to-end training, to ensure the convergence of the UAEN which lies deep inside the AE network. Simulation results demonstrated that the proposed AUD scheme achieved 3 to 5dB gain at a target activity detection error rate of $\bf{{10}^{-3}}$ compared to the state-of-the-art DL-based AUD schemes.
\end{abstract}

\begin{IEEEkeywords}
Grant-free, sparse code multiple access, active user detection, deep-learning, autoencoder, extraction network
\end{IEEEkeywords}

\section{Introduction}
\vspace*{-5pt}
In the grant-free sparse code multiple access (GF-SCMA) system, it is not explicitly known to the BS which users become active [1]-[2]. In order to identify the individual user’s activity therein, preambles are used to identify the individual users’ activity from preamble through a process known as active user detection (AUD) followed by estimation of their channels. If base station (BS) can determine which users are active through AUD, channel estimation (CE) and multi-user detection (MUD) can be performed only for the active users. Among the process in this sequence of detection, i.e., AUD, CE, and MUD, AUD is a major performance bottleneck as it involves complex combinatorial (nonlinear) binary decision process [3].

Conventionally, AUD is implemented by a compressive sensing (CS) approach, which exploits user sparsity, i.e., only a small number of devices are active among the massive number of devices [3]. However, its performance would be severely degraded when the number of contending users and their activity rate increase [4]. To tackle these challenges, deep learning (DL)-based AUD schemes have been proposed to exiploit deep neural network’s (DNN) ability of approximating a nonlinear function to solve the AUD problem [4]-[7].

In this paper, we also consider DL-based AUD while aiming at improving its performance by data-aided approach, which exploits both SCMA-based data transmission and preamble sequences at the same time. Notably, a majority of existing CS-based and DL-based AUD use only the preamble for their AUD processes. Specifically, CS-based AUD schemes [3] and DL-based AUD schemes in [4]-[5], [7] do not consider exploiting the user data, but only utilize the user-specific preamble sequences. However, user data transmitted along with the preamble for grant-free access can be exploited to improve the AUD performance without incurring additional resource overhead [1]-[2].

To differentiate AUD schemes that employ both preamble and data from the preamble-based AUD schemes, we refer to the former as data-aided AUD. Notably, there have been various types of data-aided AUD in the previous works, including DL-based AUD scheme [6] and CS-based AUD schemes [2]-[3]. However, [6] does not consider any special NN architecture or DL method that may derive user activity information from $M$-ary codewords (CWs) associated with active user. Furthermore, in CS-based AUD schemes in [2] and [3], extrinsic information on user activity needs to be extracted from their MUD block. Consequently, AUD has to be iteratively processed along with MUD, incurring high computational complexity.

To address these challenges, we propose a new DL-based data-aided AUD framework, which extracts a priori user activity information for AUD in a feed-forward manner (non-iterative fashion). This is enabled by an end-to-end training of an autoencoder (AE), which is built around a user activity extraction network (UAEN). Owing to the proposed UAEN, the computational complexity of the AUD block is independent of the length of the $M$-ary data stream considered for the AUD. Thus, an efficient design of the UAEN allows extensive increase of the number of $M$-ary CWs used to enhance its performance. In the course of end-to-end training for the proposed AE, preamble sequences can be jointly designed by considering their associated SCMA codebook (CB). Despite the potential advantages of the proposed scheme, it requires an efficient training methodology to handle the training complexity involved with joint optimization in the current framework.

Accordingly, we propose a UAEN architecture to efficiently extract a priori user activity information from the active user’s SCMA CWs. Furthermore, we also present a self-supervised pre-training scheme for the UAEN to support the improved convergence in the proposed AE architecture. It is demonstrated that when the proposed DL-based data-aided AUD architecture is sufficiently converged, it achieves 3 to 5dB gain at the target activity detection error rate (ADER) of $10^{-3}$, compared to state-of-the-art DL-based AUD schemes.

The rest of this paper is organized as follows. In Section \uppercase\expandafter{\romannumeral2}, we present a system model for GF-SCMA, including DL-based AUD. Section \uppercase\expandafter{\romannumeral3} presents a novel data-aided AUD scheme using UAEN and the proposed network structure with its training methodology. Simulation results are given and discussed in Section \uppercase\expandafter{\romannumeral4}. Finally, conclusion is made in the last section.

\vspace*{-10pt}

\section{System Model}

\subsection{Grant-free SCMA System}

We consider a single-cell GF-SCMA system with $K$ potentially transmitting users, each of them being active at each transmission opportunity. Each transmission opportunity is given as a contention transmission unit (CTU), which corresponds to a pair of CB and its associated preamble [1]. We assume that each user employs one of $J$ CBs available for data transmission. Let $\bar{\mathcal{C}}=\{{{\mathcal{C}}_{0}},{{\mathcal{C}}_{1}},\ldots ,{{\mathcal{C}}_{J-1}}\}$ denote a set of CBs with  ${{\mathcal{C}}_{j}}$ representing the $j$-th CB.

Note that the number of available CBs is limited due to the complexity of designing CB for SCMA, which grows exponentially with $J$ [8]. Therefore, the number of CBs is much smaller than the number of users, i.e., $J\ll K$. If two or more users employ the same CB, they will be collided. Therefore, a collision rate can be reduced by optimizing how the CBs are shared among all users, possibly depending on their activity probabilities. In the current design, we assume homogeneous user activity, i.e., ${{p}_{0}}={{p}_{1}}=\ldots ={{p}_{K-1}}=\bar{p}$, making a round-robin CB assignment among all users acceptable for averaging out the CB collision rate. More specifically, CB with index $\nu (k)=\bmod (k,J)$, denoted as ${{\mathcal{C}}_{\nu (k)}}$, is assigned to a user with index $k$, $k=1,2,\cdots ,K$. The CB assignment problem subject to non-homogeneous activity is beyond a scope of the current design.

In practice, the number of grant-free access users would be much larger than the number of preambles available. It implies the preamble assignment must be dynamic such that each user selects one at random out of the given set of preambles. In this case, a user identification would be only possible with additional information in the data field. When two or more users select the same preamble-CB pair, i.e., incurring preamble collision, the BS may not be able to differentiate the superposed signals. Therefore, the number of preambles associated with CBs must be sufficiently large enough to minimize preamble collision.

Our objective in this paper is to design an optimal set of the preamble sequences and a receiver structure at the same time such that ADER can be minimized. To this end, we expect cross-correlation among all pairs of preambles to be minimized subject to the preamble-CB association and the data-aided AUD in the receiver. In the current design, consider $N$ preambles available, $N\ll K$, implying that the maximum number of collisions would be limited to $N$. Furthermore, when $N > J$, i.e., CBs reused among all preambles, activities of two users with the same CBs can be still detected with the different preambles. As we aim at the preamble design for data-aided AUD, it is sufficient to consider $N$ users only, each with a unique preamble, i.e., $N=K$. In sequel, therefore, our design will be limited to a system model with $N=K$ users as illustrated in Fig. 1, where each user be indexed by $n\in \{0,1,\ldots ,N-1\}$. It is worth mentioning that $N$ in the proposed design can be sufficiently increased to minimize the preamble collision. Activity of user $n$ is indicated by a binary random variable ${{\delta }_{n}}\in \{0,1\}$ that follows Bernoulli distribution with activity probability ${{p}_{n}}$. If ${{\delta }_{n}}$ equals to one, user $n$  is active, otherwise, inactive. The activity indicators for all users constitutes a random activity vector, represented as $\bm{\delta }={{[{{\delta }_{0}},{{\delta }_{1}},\cdots ,{{\delta }_{N-1}}]}^{T}}$.

\begin{figure}[t]
\centering
\includegraphics[width=\linewidth,height=1.5in]{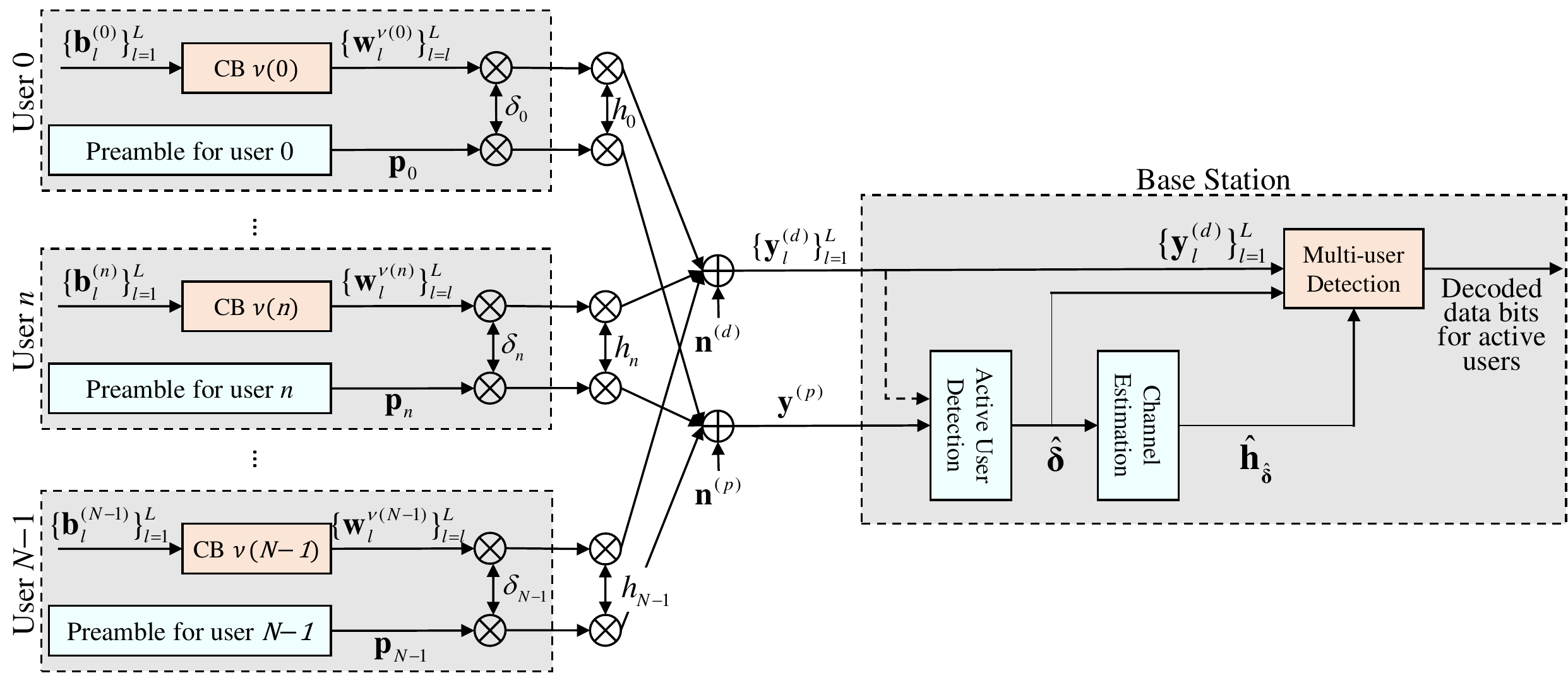}
\vspace*{-15pt}
\caption{Transceiver model for joint design of preambles and active user detection for uplink grant-free SCMA system}
\label{fig_sim}
\vspace*{-17pt}
\end{figure}

Consider an $M$-ary CW $\mathbf{c}_{m}^{\nu (n)}\in {{\mathbb{C}}^{{{K}^{(d)}}}}$ over ${{K}^{(d)}}$ orthogonal resources, constructing a pre-assigned CB ${{\mathcal{C}}_{\nu (n)}}=\{\mathbf{c}_{1}^{\nu (n)},\mathbf{c}_{2}^{\nu (n)},\ldots ,\mathbf{c}_{M}^{\nu (n)}\}$ to transmit its equally likely $M$-ary bit sequence. In order to detect the user data of grant-free uplink access, a receiver must know which CB has been employed for transmission. Towards that end, each transmission is preceded with a unique preamble, which can be used to identify its own associated CB [1]-[2]. In other words, each preamble has its own CB, which is assigned in a round-robin manner out of the CB set $\bar{\mathcal{C}}$.

Let ${{\mathbf{p}}_{n}}\in {{\mathbb{C}}^{{{K}^{(p)}}}}$ represent a preamble for user $n$ where ${{K}^{(p)}}$ denotes a dimension of the preamble. Meanwhile, a data bit sequence of active user is divided into $L$ blocks, each with ${{\log }_{2}}M$ bits. Then, each block is encoded with its own preassigned CB. Let $\mathbf{b}_{\ell }^{(n)}\in {{\mathbb{B}}^{{{\log }_{2}}M}}$ denote the $\ell $-th block of user $n$, which is encoded with a CW, denoted as $\mathbf{w}_{\ell }^{\nu (n)}\in {{\mathcal{C}}_{\nu (n)}}$. In this case, CTU for user $n$ can be represented by $\left[ {{\mathbf{p}}_{n}}\left| \mathbf{w}_{0}^{\nu (n)},\mathbf{w}_{1}^{\nu (n)}, \right.\cdots ,\mathbf{w}_{L-1}^{\nu (n)} \right]$. The simultaneous CTU’s for all active users are superposed over the same physical resources in a  contention region (time-frequency resource), i.e., preamble sequences over ${{K}^{(p)}}$ resources and the $\ell $-th CWs over $(\ell -1)\cdot {{K}^{(d)}}+1$ to $\ell \cdot {{K}^{(d)}}$ resources.

\begin{figure*}[ht]
\centering
\includegraphics[width=6.75in,height=1.65in]{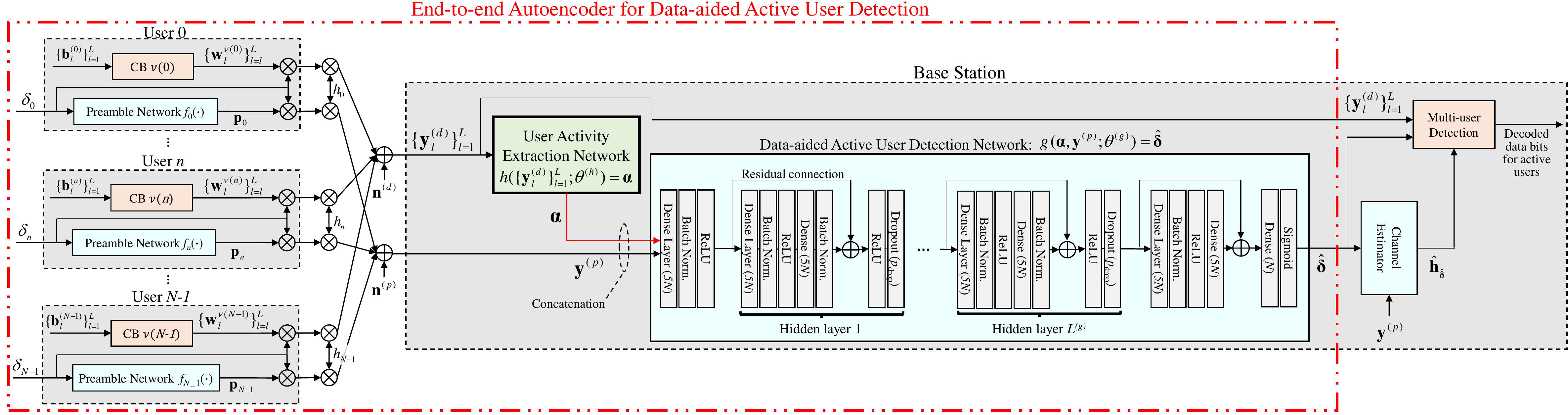}
\vspace*{-5pt}
\caption{End-to-end autoencoder architecture for proposed data-aided AUD with UAEN}
\vspace*{-15pt}
\label{fig_sim}
\end{figure*}

We assume that all CTUs are transmitted within the coherence time, therefore experiencing a flat fading channel, as in the most machine type communication  scenarios. Thus, a CTU for user $n$ is subject to a single fading channel coefficient ${{h}_{n}}$ which constitutes a system channel vector, represented as $\mathbf{h}={{[{{h}_{0}},{{h}_{1}},\ldots ,{{h}_{N-1}}]}^{T}}$. Consequently, the received signals ${{\mathbf{y}}^{(p)}}$ and $\mathbf{y}_{\ell }^{(d)}$ for the preamble and the $\ell $-th CW superposed over the given resources, respectively, are expressed as
\begin{equation}
{{\mathbf{y}}^{(p)}}=\sum\nolimits_{n=0}^{N-1}{{{\delta }_{n}}{{h}_{n}}{{\mathbf{p}}_{n}}}+{{\mathbf{n}}^{(p)}}, 
\label{eq}
\end{equation}
\begin{equation}
\mathbf{y}_{\ell }^{(d)}=\sum\nolimits_{n=0}^{N-1}{{{\delta }_{n}}{{h}_{n}}\mathbf{w}_{\ell }^{\nu (n)}}+\mathbf{n}_{\ell }^{(d)}, \ell =1,2,\cdots ,L.
\label{eq}
\end{equation}

In Fig. 1, AUD is performed only using the superposed preamble ${{\mathbf{y}}^{(p)}}$ to detect an activity vector, represented as $\bm{\hat{\delta }}={{[{{\hat{\delta }}_{0}},{{\hat{\delta }}_{1}},\ldots ,{{\hat{\delta }}_{N-1}}]}^{T}}$. After the AUD process, channel coefficients for the active users are estimated by a CE process, e.g., using a linear minimum mean square (LMMSE) estimator. Let ${{\hat{h}}_{n}}$ represent an estimated channel of active user $n$. Then, an estimated system channel vector can be represented as ${{\mathbf{\hat{h}}}_{{\bm{\hat{\delta }}}}}={{[{{\delta }_{0}}{{\hat{h}}_{0}},{{\delta }_{1}}{{\hat{h}}_{1}},\ldots ,{{\delta }_{N-1}}{{\hat{h}}_{N-1}}]}^{T}}$. Finally, active users’ transmitted bit sequences is detected by a BS performs MUD, e.g., the message passing algorithm (MPA)-based MUD [8], using $\bm{\hat{\delta }}$, ${{\mathbf{\hat{h}}}_{{\bm{\hat{\delta }}}}}$, and the received data stream $\{\mathbf{y}_{l}^{(d)}\}_{l=1}^{L}$. The objective of this paper is to design the preambles $\{{{\mathbf{p}}_{n}}\}_{n=0}^{N-1}$ and an AUD scheme using DL-based joint optimization for the GF-SCMA system.

\subsection{Deep learning-based Active User Detection}
The DL-based AUD schemes have been demonstrated to successfully achieve superior performance over conventional CS-based AUD schemes [4]-[7]. The objective of these DL-based AUDs is to optimize an AUD network (AUDN), denoted as $g(\cdot )$, in terms of its trainable parameters. More specifically, the AUDN must be designed to minimize the ADER, which can be formulated as the following optimization problem:
\begin{equation}
{{g}^{*}}=\underset{g}{\mathop{\arg \min }}\,{{\left\| \bm{\delta }-g(\cdot ;{{\mathbf{\theta }}^{(g)}}) \right\|}_{0}},
\label{eq}
\end{equation}
where ${{\bm{\theta }}^{(g)}}$ is a vector of weight and bias AUDN and ${{\left\| \cdot  \right\|}_{0}}$ denotes L0-norm. Depending on whether it is preamble-based AUD [4]-[5], [7] or data-aided AUD [6], AUDN is given as $g(\cdot ;{{\bm{\theta }}^{(g)}})={{g}_{1}}({{\mathbf{y}}^{(p)}};{{\bm{\theta }}^{({{g}_{1}})}})$ or $g(\cdot ;{{\bm{\theta }}^{(g)}})={{g}_{2}}({{\mathbf{y}}^{(p)}},\{\mathbf{y}_{l}^{(d)}\}_{l=1}^{L};{{\bm{\theta }}^{({{g}_{2}})}})$, respectively, where ${{\bm{\theta }}^{({{g}_{k}})}}$ is a vector of weight and bias of ${{g}_{k}}(\cdot )$. Both ${{g}_{1}}({{\mathbf{y}}^{(p)}};{{\bm{\theta }}^{({{g}_{1}})}})$ and ${{g}_{2}}({{\mathbf{y}}^{(p)}},\{\mathbf{y}_{l}^{(d)}\}_{l=1}^{L};{{\bm{\theta }}^{({{g}_{2}})}})$ are trained to detect user activity $\bm{\hat{\delta }}$, i.e., $\bm{\hat{\delta }}={{g}_{1}}({{\mathbf{y}}^{(p)}};{{\bm{\theta }}^{({{g}_{1}})}})$ and $\bm{\hat{\delta }}={{g}_{2}}({{\mathbf{y}}^{(p)}},\{\mathbf{y}_{l}^{(d)}\}_{l=1}^{L};{{\bm{\theta }}^{({{g}_{2}})}})$.

Since ${{\mathbf{y}}^{(p)}}$ and $\{\mathbf{y}_{\ell }^{(d)}\}_{\ell =1}^{L}$ over the same resources are transmitted from the same active users, their joint distributions can be exploited to improve the AUDN performance. However, if the two heterogeneous measurements are jointly processed in a single network, there are too many possible measurements to consider which makes a training process to optimize the AUDN difficult and complex. In order to handle the concerned training complexity, another entity, referred to as user activity extraction network (UAEN), is introduced. A notion of the extraction network was recently introduced for pilotless communication to extract channel information from the $M$-ary data stream for end-to-end training of an AE [9]. However, the proposed UAEN processes $\{\mathbf{y}_{\ell }^{(d)}\}_{\ell =1}^{L}$ to extract an additional feature of user activity. The new feature will be subsequently exploited by AUDN, along with ${{\mathbf{y}}^{(p)}}$.

\section{Proposed Joint Design for Data-aided Active User Detection}
\subsection{Overall End-to-end AE Architecture}
In this section, we present a proposed AE architecture with UAEN and its training methodology for the data-aided AUD framework with a design model in Fig. 1. As illustrated in Fig. 2, overall architecture of the proposed end-to-end AE deals with preamble generation network (PGN), denoted as $\{{{f}_{n}}(\cdot )\}_{n=0}^{N-1}$, for each user, and data-aided AUDN and UAEN, denoted as $g(\cdot )$, and $h(\cdot )$, respectively in the BS. It aims at joint optimization of all networks, $\{{{f}_{n}}(\cdot )\}_{n=0}^{N-1}$, $g(\cdot )$, and $h(\cdot )$.

Our design objective is to determine $N$ unique preamble sequences which will be shared among the multiple users. To this end, we assume that each user is assigned with a unique preamble, as shown in Fig. 2. Furthermore, SCMA CBs are uniformly shared among users, i.e., by allocating $J$ CBs to $N$ users in a round-robin manner. The SCMA CBs herein are optimized independently, e.g., using the conventional SCMA CB in [8]. Then, the AE in Fig. 2 allows for generating $N$ unique preamble sequences, each associated with one of $J$ CBs, subject to a random activity vector. These sequences must be determined such that the cross-correlation among them is minimized. In other words, we focus on the preamble design only with the end-to-end configuration in Fig. 2, i.e., without considering a collision performance as in [4]-[7].

A PGN ${{f}_{n}}(\cdot )$ is an NN with the weight and bias of ${{\mathbf{\theta }}^{(f)}}$ that provides a preamble sequence for active user $n$, i.e., when ${{\delta }_{n}}$ = 1; otherwise, it gives all-zero sequence, indicating inactive transmission. Then, the output of PGN can be represented as
\begin{equation}
{{f}_{n}}({{\delta }_{n}};{{\bm{\theta }}^{(f)}})=\left\{ \begin{matrix}
   {{\mathbf{p}}_{n}},\text{  }{{\delta }_{n}}=1,  \\
   \mathbf{0},\text{   }{{\delta }_{n}}=0,  \\
\end{matrix} \right.
\label{eq}
\end{equation}
As in [5], ${{f}_{n}}(\cdot )$ is designed as a single dense layer with dimension of $2{{K}^{(p)}}$. The preamble is normalized as ${{\left\| {{\mathbf{p}}_{n}} \right\|}_{2}}=1$.

Meanwhile, the UAEN aims at producing a priori probabilities of user activity for individual users, denoted as a vector ${\bm{\alpha }}=[{{\alpha }_{0}},{{\alpha }_{1}},\cdots ,{{\alpha }_{N-1}}]$, where ${{\alpha }_{n}}$ represents a probability that user $n$ is active, given $\{\mathbf{y}_{\ell }^{(d)}\}_{\ell =1}^{L}$. More specifically, the UAEN is represented as
\begin{equation}
h(\{\mathbf{y}_{l}^{(d)}\}_{l=1}^{L};{{\bm{\theta }}^{(h)}})=\bm{\alpha },
\label{eq}
\end{equation}
where ${{\mathbf{\theta }}^{(h)}}$ is a vector of weight and bias in the network. The detailed design approach is discussed in the sequel.

As $\bm{\alpha }$ is exploited as a priori information for our proposed data-aided AUD, the proposed AUDN can be represented as
\begin{equation}
g(\bm{\alpha },{{\mathbf{y}}^{(p)}};{{\bm{\theta }}^{(g)}})=\bm{\hat{\delta }}
\label{eq}
\end{equation}
where ${{\mathbf{\theta }}^{(g)}}$ is a vector of weight and bias in the proposed data-aided AUD with UAEN. The user activity is finally determined by a threshold criterion, i.e., user n is declared active if ${{\hat{\delta }}_{n}}>\gamma $, where $\gamma $ denotes an activity threshold [5]. A serially concatenated vector of $\bm{\alpha }$ and ${{\mathbf{y}}^{(p)}}$ is taken as an input to the data-aided AUD. As $\bm{\alpha }$ is user activity information extracted from the $\{\mathbf{y}_{\ell }^{(d)}\}_{\ell =1}^{L}$, the proposed data-aided AUD is expected to improve the performance through activity information jointly extracted from both data and preamble.

The detailed structure of AUDN is given as deep residual network with ${{L}^{(g)}}$ hidden layers as in conventional DL-based AUD schemes [4]-[5], [7]. As illustrated in Fig. 3, each hidden layer is a combination of dense layers with $5N$ nodes, batch normalization, rectified linear unit (ReLU) activation function, and dropout layer with dropout probability ${{p}_{drop}}$ as illustrated in Fig. 2. The residual connections, ReLU activation function, and batch normalization inside each hidden layer are essential for improving the convergence performance while the dropout layer facilitates generalization of AUD [4], [7]. Meanwhile, the AUDN can be implemented with long short-term memory network as in [6]. However, to fairly compare only data-aided performance gain over the conventional AUDN, implementation of deep residual network-based AUDN is considered throughout the paper.

Ideally, it is possible to jointly train the PGN, AUDN, and UAEN only by end-to-end training through the following binary cross-entropy loss function:
\begin{equation}
\resizebox{.90 \columnwidth}{!} 
{
${{\mathcal{L}}_{E2E}}(\bm{\delta },\bm{\hat{\delta }})=-\sum\limits_{n=0}^{N-1}{({{\delta }_{n}}\log ({{{\hat{\delta }}}_{n}})+(1-{{\delta }_{n}})\log (1-{{{\hat{\delta }}}_{n}}))}.$

}
\label{eq}
\end{equation}
Note that the loss function in (7) is also considered in [4]-[7]. However, since the proposed UAEN is very deep inside the AE, the gradient reached for the UAEN using (7) will be vanished during the backpropagation process, deteriorating the convergence performance. To solve this problem, a pre-training scheme and a special type of network architecture are proposed for UAEN in the following subsections.

\subsection{Self-supervised Pre-training Scheme for UAEN}
We note that another network employed to extract a priori user activity information makes the AE architecture much deeper than the existing DL-based AUD in [4]-[7]. The deeper network might suffer from a vanishing gradient problem if a training methodology is similar to that in [4]-[7]. Notably, pre-training is considered as one of the powerful steps to solve the problem while training a deep network along with weight initialization, activation function, and batch normalization layer [10]. To pre-train the UAEN, we employ self-supervised learning for the UAEN in such a way that its output $\bm{\alpha }$ becomes close to the activity vector $\bm{\delta }$. Thus, the following cross-entropy loss function can be used for pre-training: 
\begin{equation}
\resizebox{.90 \columnwidth}{!} 
{

${{\mathcal{L}}_{PT}}(\bm{\delta },\bm{\alpha })=-\sum\limits_{n=0}^{N-1}{({{\delta }_{n}}\log ({{\alpha }_{n}})+(1-{{\delta }_{n}})\log (1-{{\alpha }_{n}}))}.$

}
\label{eq}
\end{equation}
However, by only using (8), the UAEN is pre-trained without any consideration on the AUDN. Thus, it should be fine-tuned by jointly training both AUDN and UAEN in an end-to-end manner with (7). The end-to-end learning process renders $\bm{\alpha }$ to be a priori user activity information when jointly considering UAEN and AUDN. Toward this end, the pre-training (Step 1) and end-to-end training (Step 2) stages are combined to form a two-step training approach, which is detailed in the sequel.

In Step 1, the UAEN is trained at a learning rate $\eta $ over $T_{1}^{(1)}$ epochs and then, trained at a reduced learning rate $\eta /10$ over $T_{2}^{(1)}$ epochs. The two different levels of learning rate allow for sufficient convergence in the pre-training stage. Similarly in Step 2, the AE is trained with reducing the learning rates over different periods of epochs. Consider a training process over the multiple periods of epochs, e.g., $(T_{1}^{(2)},T_{2}^{(2)},\cdots ,T_{Q}^{(2)})$, where $Q$ denotes the number of different periods, each with $T_{n}^{(2)}$ epochs, $n=1,2,\cdots ,Q$. Let ${{\eta }_{n}}$ denote a learning rate for the n-th period of training in Step 2. Then, that for the $n$-th period is set to ${{\eta }_{n}}={{\eta }_{0}}/{{10}^{n-1}}$, where ${{\eta }_{0}}$ denotes an initial learning rate.

\subsection{Design of UAEN}
To efficiently extract a priori activity vector ${\bm{\alpha }}$ from the $\{\mathbf{y}_{l}^{(d)}\}_{l=1}^{L}$, we propose a special type of UAEN structure as shown in Fig. 3. To justify the proposed architecture, the relationship between $\{\mathbf{y}_{l}^{(d)}\}_{l=1}^{L}$ and activity vector $\bm{\delta}$ should be reflected to the NN structure. For example, ${{N}_{a}}$ active users lead to ${{M}^{{{N}_{a}}}}$ possible realizations of $\mathbf{y}_{l}^{(d)}$. It implies that $\mathbf{y}_{l}^{(d)}$ and $\mathbf{y}_{l+1}^{(d)}$ can be completely different from each other even subject to the same activity vector $\bm{\delta }$. Moreover, $L$ consecutive superposed CWs, i.e., $\{\mathbf{y}_{l}^{(d)}\}_{l=1}^{L}$, would lead to ${{({{M}^{{{N}_{a}}}})}^{L}}$ different realizations for the given activity vector $\bm{\delta }$. Thus, joint processing of $\{\mathbf{y}_{l}^{(d)}\}_{l=1}^{L}$ in UAEN generates more diverse features to consider for extracting user activity as $L$ increases. However, joint processing of $\{\mathbf{y}_{l}^{(d)}\}_{l=1}^{L}$ involves enlarging the UAEN, incurring a vanishing gradient problem while involving a high computational complexity. To handle these problems, the number of different data samples that the UAEN should jointly consider can be reduced by processing $\{\mathbf{y}_{l}^{(d)}\}_{l=1}^{L}$ via independent smaller networks and then, combining their output later. As shown in Fig. 3, therefore, we independently process $\{\mathbf{y}_{l}^{(d)}\}_{l=1}^{L}$ using 1-D convolution layers to provide intermediate activity vectors, then combining them together to generate $\bm{\alpha }$ using multiple dense layers.

The detailed structure of the proposed UAEN can be determined by the three hyperparameters, $N_{\operatorname{ke}\text{rnel}}^{(1)}$, $N_{\operatorname{ke}\text{rnel}}^{(2)}$, and ${{L}^{(h)}}$. We employ two 1-D convolutional layers to produce an intermediate activity vector. The first convolutional layer is made of $N_{\operatorname{ke}\text{rnel}}^{(1)}$ kernels with a size of $2\cdot {{K}^{(d)}}$. Let ${{\bm{\alpha }''}_{l}}\in {{\mathbb{R}}^{N_{\text{kernel}}^{(1)}}}$ represent its output for $\mathbf{y}_{l}^{(d)}$. In order to build an acceptable size of dense network for combining the intermediate activity vectors, we consider another convolutional layer that can reduce the number of input parameters. It is made of $N_{\operatorname{ke}\text{rnel}}^{(2)}$ kernels with a size of $N_{\operatorname{ke}\text{rnel}}^{(1)}$ ($N_{\operatorname{ke}\text{rnel}}^{(1)}>N_{\operatorname{ke}\text{rnel}}^{(2)}$). Let ${{\bm{\alpha }'}_{l}}\in {{\mathbb{R}}^{N_{\text{kernel}}^{(2)}}}$ represent its output for $\mathbf{y}_{l}^{(d)}$, forming the intermediate activity vector. After constructing $\{{\bm{\alpha }'}_{l}\}_{l=1}^{L}$, ${{L}^{(h)}}$ dense layers are applied to the UAEN with $({{L}^{(h)}}-i+1)\cdot N$ nodes in the $i$-th layer, $i=1,2,\cdots ,{{L}^{(h)}}$. Consequently, multiple dense layers are adopted to combine $\{{\bm{\alpha }'}_{l}\}_{l=1}^{L}$ for producing the activity vector $\bm{\alpha }\in {{\mathbb{R}}^{N}}$. Note that batch normalization and ReLU activation function are used after every convolutional and dense layers.

\begin{figure}[t]
\centering
\includegraphics[width=3in,height=2.1in]{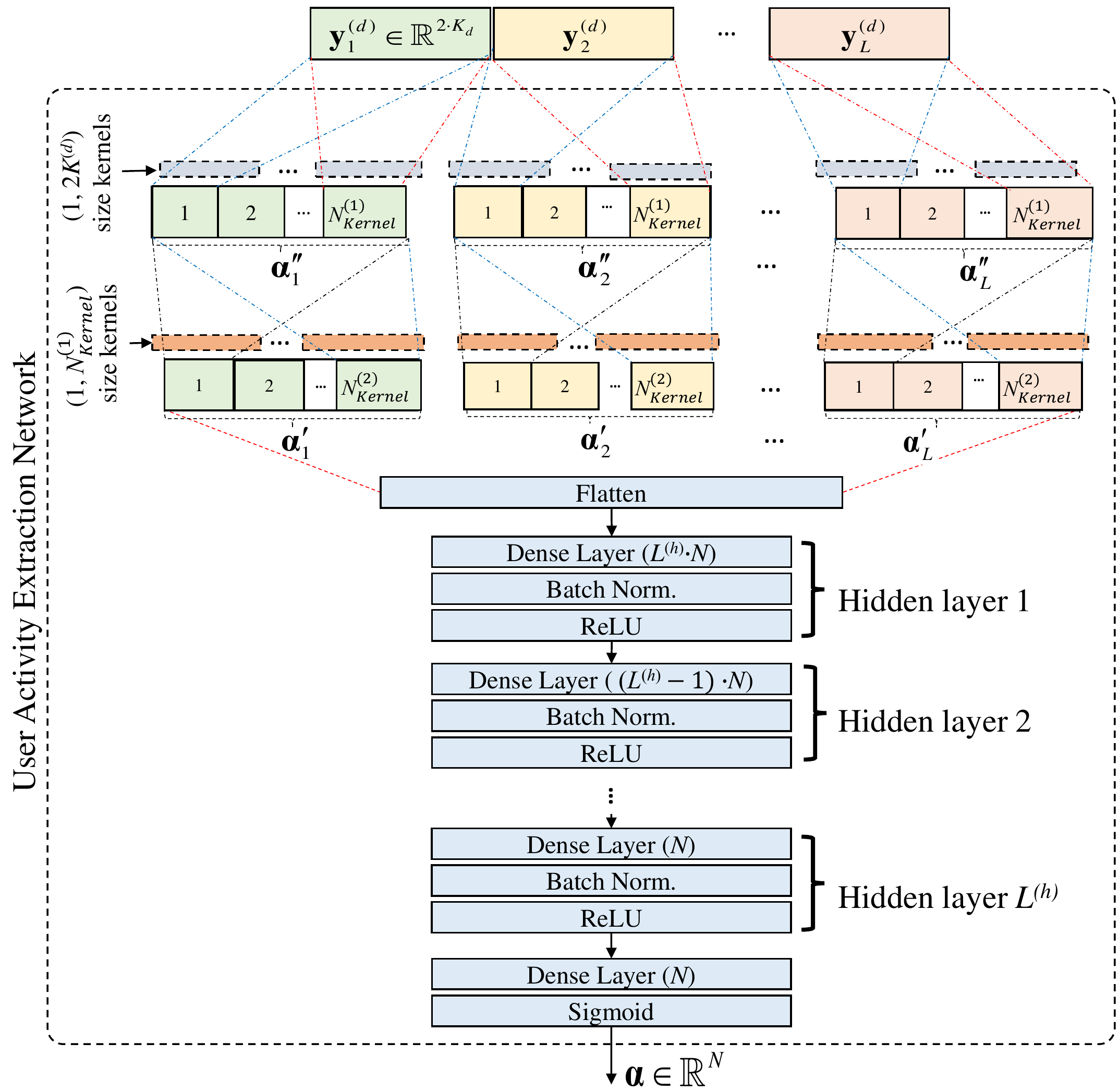}
\vspace*{-5pt}
\caption{Detailed structure of user activity extraction network}
\label{fig_sim}
\vspace*{-15pt}
\end{figure}

\section{Simulations and Discussions}
\subsection{Simulation Setup}
We consider a GF-SCMA system with $N=64$ users, in which a well-known SCMA CB in [8] is used, i.e., $J=6$, ${{K}^{(d)}}=4$ and $M=4$. The length of preamble sequence is given as ${{K}^{(p)}}=16$. We assume homogeneous activity probability is set to either $\bar{p}=0.05$ or $\bar{p}=0.1$ in the current investigation. The AUD performance is mainly governed by the cross-correlation between any two active preambles, which is a channel-agnostic metric. It can be well optimized through the offline training of AE under an AWGN channel [5]. Toward this end, we consider the AWGN channel model with a range of signal-to-noise ratio (SNR), which is uniformly distributed between 15dB and 20dB.

Data samples are constructed by activity vectors $\bm{\delta }$ for the given $\bar{p}$. In the first step, we generate 250,000 random activity samples for training data while 480,000 samples are used in the second step. In both cases, 60,000 samples are used for validation data. Finally, 60,000 samples are used for evaluating the final ADER performance. Note that ADER is evaluated by considering the number of users subject to misdetection and false alarms of AUD among all potential active users. We employ an ADAM optimizer with a batch size of 20 and a threshold set to $\gamma =0.4$. [5] The detailed hyperparameters for the proposed scheme are summarized in Table 1. Note that the value of $N_{\text{kernel}}^{(1)}$ and $N_{\text{kernel}}^{(2)}$ are increased for the higher activity probability. In these cases, the number of trainable parameters in the UAEN are only 6.33\% and 20.17\% of those of AUDN for $\bar{p}=0.05$ and $\bar{p}=0.1$, respectively.

\begin{table}[htbp]
\centering
\caption{Summary of Hyperparameters}
\vspace*{-12pt}

\scalebox{0.85}{

\begin{tabular}{M{0.15\columnwidth-2\tabcolsep}|M{0.38\columnwidth-2\tabcolsep}|M{0.2\columnwidth-2\tabcolsep}|M{0.27\columnwidth-2\tabcolsep}} 
\noalign{\smallskip}\noalign{\smallskip}\hline\hline
\multicolumn{2}{c|}{\textbf{Parameter}} & \textbf{Notation} & \textbf{Value} \\
\hline
\multirow{9}{*}{\makecell{UAEN}}
						& Size of the first kernel & $N_{\text{kernel}}^{(1)}$ & 256 ($\bar{p}=0.05$) or 512 ($\bar{p}=0.1$)  \\ \cline{2-4} 
						& Size of the second kernel & $N_{\text{kernel}}^{(2)}$ & 32 ($\bar{p}=0.05$) or 64 ($\bar{p}=0.1$)  \\ \cline{2-4} 
						& Number of hidden layers & ${{L}^{(h)}}$ & 3  \\ \cline{2-4} 
						& Number of epochs in the first training steps & $T_{1}^{(1)}$ & 15 \\ \cline{2-4} 
						& Number of epochs in the second training steps & $T_{2}^{(1)}$ & 10  \\ \cline{2-4} 
						& Learning rate & $\eta$ & 0.01  \\ \hline

\multirow{6}{*}{\makecell{AUDN}}	
						& Dropout probability & ${{p}_{drop}}$ & 0.1  \\ \cline{2-4} 
						& Number of hidden layers & ${{L}^{(g)}}$ & 10  \\ \cline{2-4} 
						& Initial learning rate & ${{\eta }_{0}}$ & 0.01  \\ \cline{2-4} 
						& Number of training steps & $Q$ & 4 \\ \cline{2-4}
						& Number of epochs in the different training steps & $\{T_{n}^{(2)}\}_{n=1}^{4}$ & 10  \\ 
\hline
\hline
\end{tabular}

}

\vspace*{-10pt}
\end{table}

\begin{figure}[t]
\centering
\includegraphics[width=2.6in,height=1.845in]{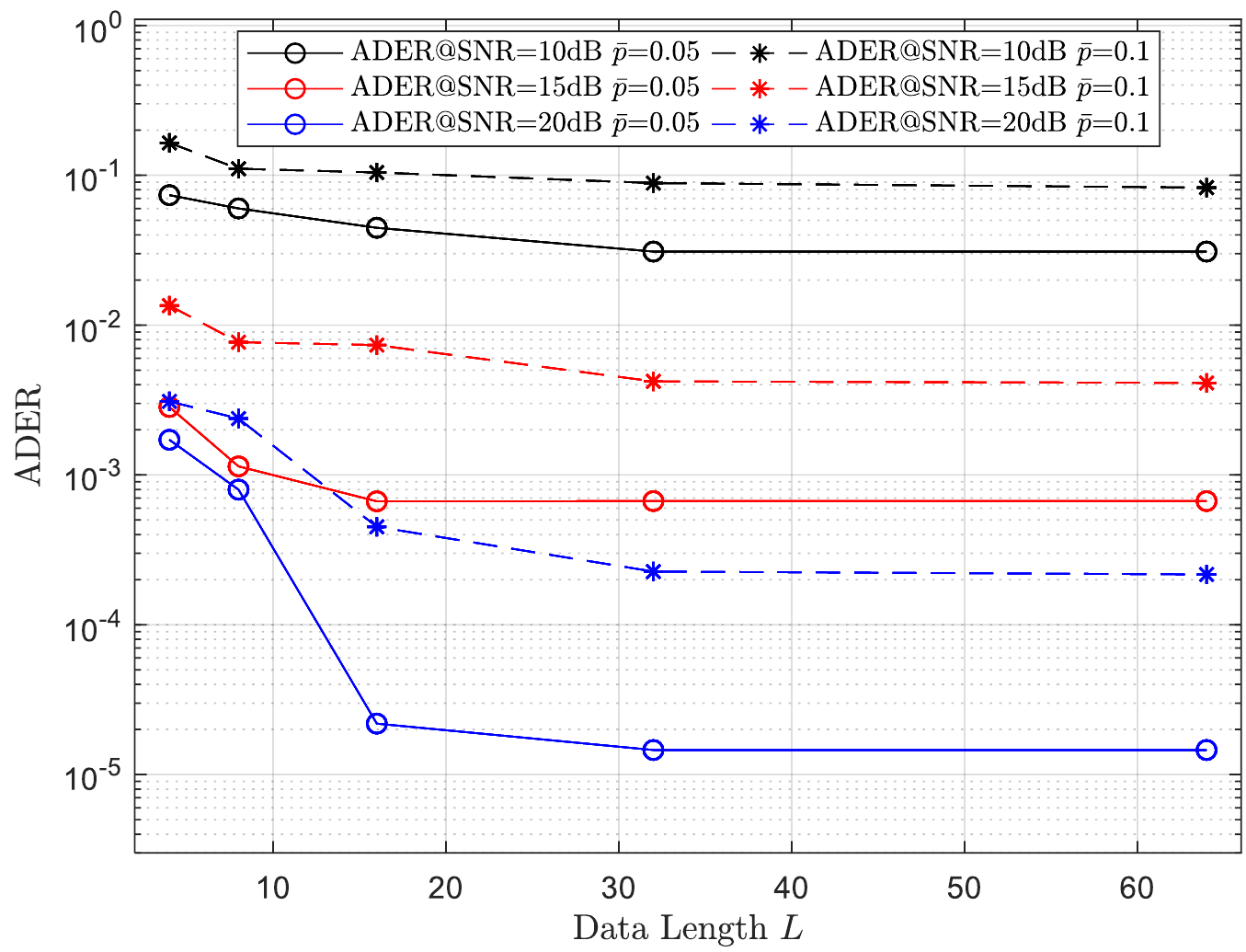}
\vspace*{-5pt}
\caption{ADER on different data length}
\label{fig_sim}
\vspace*{-15pt}
\end{figure}

\subsection{Simulation Results}
Fig. 4 presents ADER performance of the proposed data-aided AUD scheme as varying the data length with $L=4$, 8, 16, 32 and 64. The solid and dotted lines in the figure represents the ADER for $\bar{p}=0.05$ and $\bar{p}=0.1$, respectively. If the data length $L$ is too small, it is not possible to extract sufficient a priori information for AUDN. It is clear that the ADER drops with increasing the data length $L$. However, if $L$ is too large, the computational complexity of UAEN increases. Therefore, it is important to select appropriate value of $L$, depending on the system parameters to trade-off the ADER performance with the UAEN complexity. In our current analysis, we set the data length to $L$=16 and $L$=32 for $\bar{p}=0.05$ and $\bar{p}=0.1$, respectively.

Fig. 5 present the ADER performance of DL-based AUD for $\bar{p}=0.05$ with the different NN architectures as SNR varies. They include the conventional schemes, such as preamble-based AUD [5] and DL-based data-aided AUD [6], along with the proposed DL-based data aided AUD using UAEN. We also consider various structures of UAENs, while maintain the same level of complexity in terms of the number of trainable parameters. In particular, we consider a fully connected (FC) network with ${{L}^{(h)}}=6$ hidden layers for the UAEN, rather than 1-D convolutional layers. Its performance is evaluated with and without pre-training. When only preamble is exploited for the AUDN as in [5], its performance serves as a lower bound for the data-aid AUD schemes. Notably, data-aided AUD with FC UAEN without pre-training performs worse than one without UAEN. It is attributed to the unoptimized convergence performance of the UAEN, implying that two-step training is essential. It is shown that data-aided AUD with FC UAEN outperforms the conventional data-aided AUD only after two-step training is applied to FC UAEN. Finally, when both the two-step training and proposed UAEN structure is employed, it gives 3dB to 5dB gain at ADER of ${10}^{-3}$, compared to [5] and [6], respectively.

In Fig. 6, the ADER performance between different training schemes for the proposed data-aided AUD is presented. When UAEN is pre-trained using loss function (8) and fix the network during the end-to-end training using (7), i.e., without UAEN finetuning, it shows data-aid performance gain compared to preamble-based AUD. However, it gives worse performance than the ADER without any pre-training. This implies that joint training of AUDN and UAEN is more critically improve the ADER performance than the existence of pre-training scheme. Consequently, when both pre-training and end-to-end training is performed with UAEN fine tuning, it gives the most optimized ADER performance due to the improved convergence performance.

\section{Conclusion}
This paper proposes a new type of DL-based data-aided AUD scheme using an UAEN in the GF-SCMA system. The proposed AUD scheme can benefit from the a priori user activity information with marginal increment of the complexity due to the UAEN extracting user activity information from superposed SCMA CWs in a feed-forward manner. Our simulation results verify that the proposed AUD scheme outperforms the existing state-of-the-art approaches when the proposed UAEN structure and pre-training method were applied together. As a future work, the overall system must be evaluated in terms of data detection performance, which requires an online training methodology for both AUDN and UAEN to learn the user’s channel as a random-access signature along with CBs and their associated preambles.

\begin{figure}[t]
\centering
\includegraphics[width=2.65in,height=1.85in]{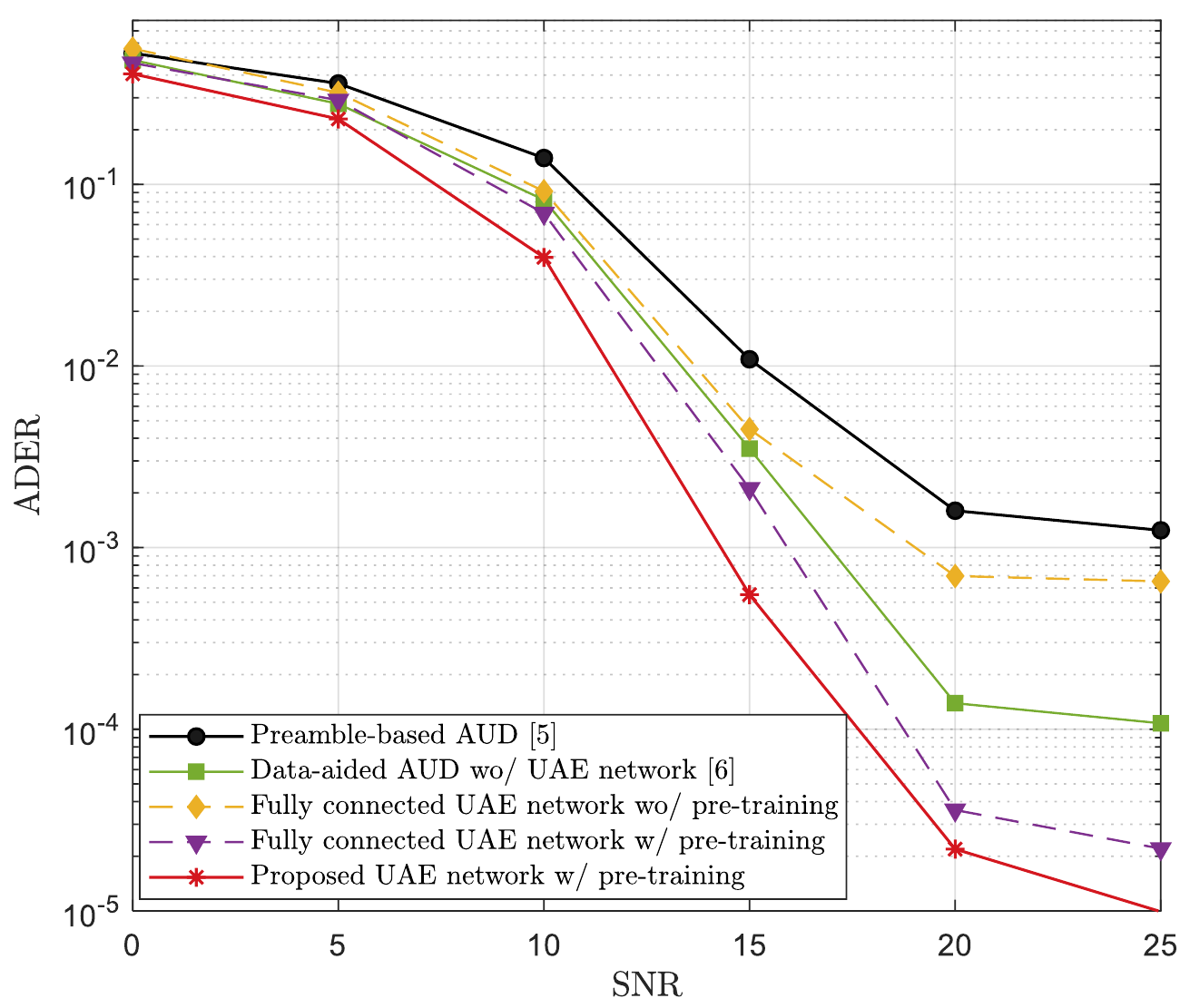}
\vspace*{-6pt}
\caption{ADER with different data-aided AUD structures}
\label{fig_sim}
\vspace*{-15pt}
\end{figure}

\begin{figure}[t]
\centering
\includegraphics[width=2.65in,height=1.85in]{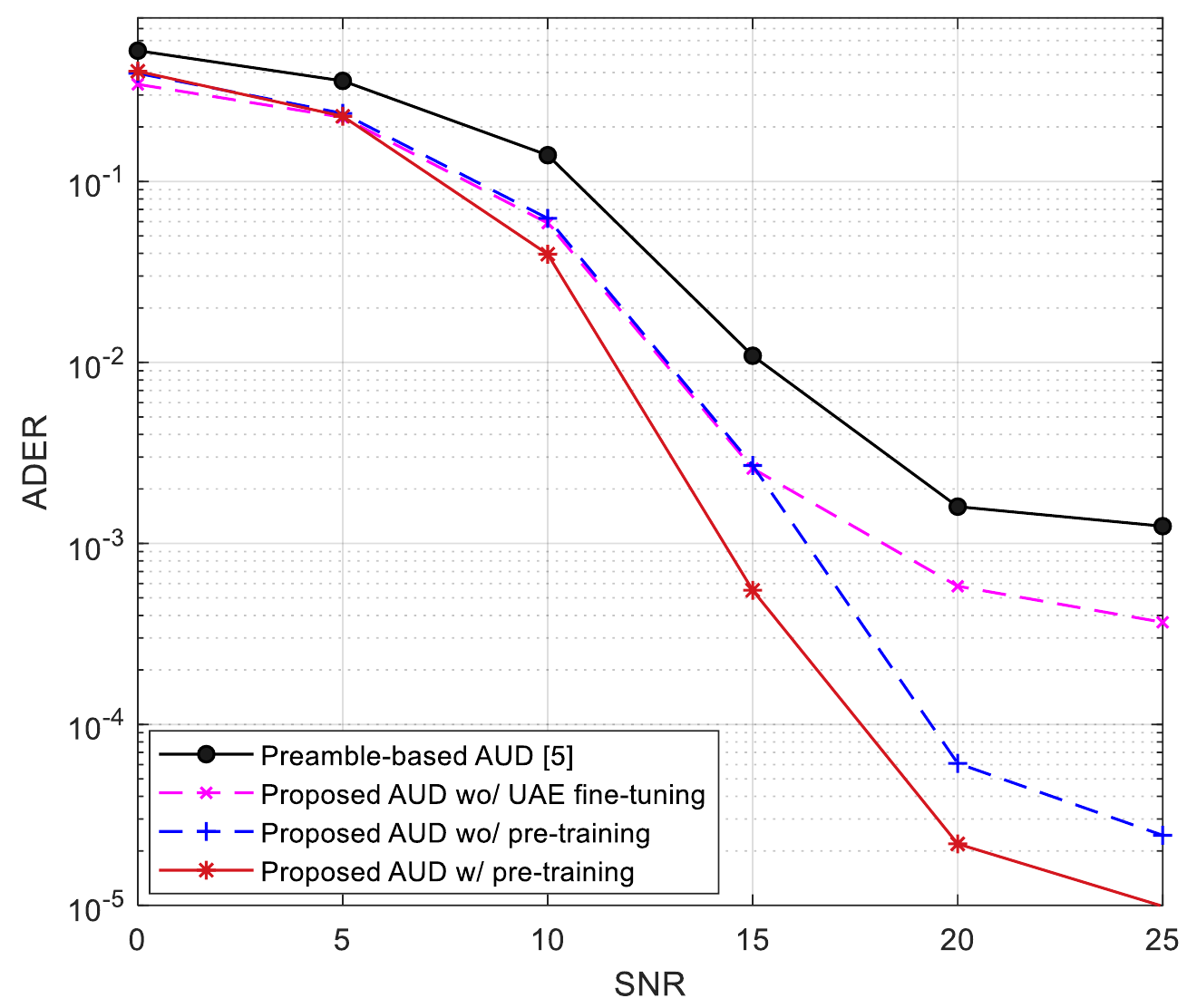}
\vspace*{-6pt}
\caption{ADER with different training schemes}
\label{fig_sim}
\vspace*{-15pt}
\end{figure}

\section*{Acknowledgment}
\vspace*{-5pt}
This work was supported in part by Institute of Information \& communications Technology Planning \& Evaluation (IITP) grant funded by the Korea government (MSIT) (No.2021-0-00467, Intelligent 6G Wireless Access System) and in part by the National Research Foundation of Korea (NRF) grant funded by the Korea government (MSIT) (No.2020R1A2C100998413).

\end{document}